\documentclass[a4paper,doc,natbib]{apa6}

\usepackage[english]{babel}
\usepackage[utf8x]{inputenc}
\usepackage{amsmath}
\usepackage{graphicx}
\usepackage{xcolor}
\usepackage{caption}
\usepackage{ragged2e}

\usepackage{euscript}
\def\pp{{\EuScript P}} 
\def\nn{{\EuScript N}} 

\usepackage{listings}
\title{Bayesian Paired Comparison with the bpcs package}
\shorttitle{Bayesian Paired Comparison with the bpcs package}
\twoauthors{David Issa Mattos}{Érika Martins Silva Ramos}
\twoaffiliations{Department of Computer Science and Engineering, Chalmers University of Technology, Gothenburg, Sweden}{Department of Psychology, University of Gothenburg, Gothenburg, Sweden}

\abstract{
\justifying
This article introduces the \texttt{bpcs} R package (Bayesian Paired Comparison in Stan) and the statistical models implemented in the package. This package aims to facilitate the use of Bayesian models for paired comparison data in behavioral research. Bayesian analysis of paired comparison data allows parameter estimation even in conditions where the maximum likelihood does not exist, allows easy extension of paired comparison models, provide straightforward  interpretation of the results with credible intervals, have better control of type I error, have more robust evidence towards the null hypothesis, allows propagation of uncertainties, includes prior information, and perform well when handling models with many parameters and latent variables. The \texttt{bpcs} package provides a consistent interface for R users and several functions to evaluate the posterior distribution of all parameters, to estimate the posterior distribution of any contest between items, and to obtain the posterior distribution of the ranks. Three reanalyses of recent studies that used the frequentist Bradley-Terry model are presented. These reanalyses are conducted with the Bayesian models of the \texttt{bpcs} package, and all the code used to fit the models, generate the figures, and the tables are available in the online appendix.}

\begin{document}
\authornote{Funding: This work was partially supported by the Wallenberg Artificial Intelligence, Autonomous Systems and Software Program (WASP) funded by Knut and Alice Wallenberg Foundation. 

The authors would like to thank L. Hopper for revising the results of the second reanalysis and G. Marton for revising the results of the third reanalysis.
} 

\maketitle

\section{Introduction}

Paired comparison data analysis arises in several contexts, such as selecting preferences or ranking items \citep{cattelan2012models}. For example, a person might be presented with questions such as ``Which brand of pizza do you prefer?'' and needs to choose between pairs, such as ``Tombstone or DiGiorno?'' or ``DiGiorno or Freschetta?''  \citep{luckett2020estimates}. The most common modeling technique for paired comparisons and the focus of this article is the Bradley-Terry model and its extensions \citep{bradley1952rank}. 

Ordinal scales can also be used to assess preferences, but they may lead to several difficulties. For example, participants may struggle to use the scale correctly \citep{coetzee1996use, petrou2003methodological}, they may try to self-monitor their answers \citep{kreitchmann2019controlling, hontangas2015comparing}, and for specific samples, the scales may not even be effective, such as in animal behavior studies, studies with young children, people with low literacy, or when respondents are using their second language to answer the scales \citep{luckett2020estimates, hopper2019assessment,huskisson2020using}. 






Paired comparisons (also called forced-choice assessments) may be stimulus-centric or person-centric. Psychometric research has developed many model applications for behavioral choice theories that consider the individual differences in psychological attributes (person-centric). These models employ item response theory (IRT) methods, such as the multidimensional pairwise comparison (MPC) \citep{wang2017item} and the multi-unidimensional pairwise preference two-parameter logistic (MUPP-2PL) model \citep{morillo2016dominance}. This line of research focuses on person-centric assessments, which have been vastly used in personality and attitudinal research \citep{brown2016item}.

In the context of this article and the context of applied psychological and behavioral research, the stimulus-centric approach is the one that fits better, and it is in this context that the Bradley-Terry model could be most useful. Examples are the preferences of natural landscapes \citep{hagerhall2018humans}, preferences for stimulus \citep{chien2012or}, for food \citep{luckett2020estimates, coetzee1996use}, analysis of animal dominance \citep{abalos2016role, bush2016lizards, miller2017fighting} and in the use of pharmacological medication \citep{meid2016longitudinal}.

Many models have been developed to extend the Bradley-Terry model. For example, to include ties in the comparisons \citep{davidson1970extending}, to address the problem of ordering items' presentation \citep{davidson1977extending}, to compensate for dependency on the data and subject-specific predictors \citep{bockenholt2001hierarchical}, or to add explanatory variables to the items \citep{springall1973response}. 

Although efforts have been made in statistical computing to provide more accurate standard errors and p-values estimates for the analyses with the Bradley-Terry model \citep{turner2012bradley, cattelan2012models}, most of these works have been focusing on providing software for the frequentist Bradley-Terry model based on the maximum likelihood \citep{turner2012bradley, hatzinger2012prefmod, gnm}. There are no comprehensive software packages that implement a Bayesian version of the Bradley-Terry model and its many extensions.

Therefore, this article proposes a Bayesian perspective to work with the Bradley-Terry model. Bayesian data analysis can provide advantages to frequentist estimation of paired comparison data. For example, it can provide parameter estimates without the need of specifying a maximum likelihood (allowing to incorporate extensions easily) and in problems where the maximum likelihood does not exist or leads to undetermined  probabilities \citep{ford1957solution, phelan2017hierarchical, butler2004existence}. Additionally, it provides better control of type I error \citep{kelter2020analysis}, provides more robust evidence towards the null hypothesis \citep{kruschke2013bayesian}, handles models with many parameters and latent variables \citep{kucukelbir2015automatic, carpenter2017stan} and allows a probabilistic interpretation of parameter intervals, as opposed to the repeated sampling interpretation \citep{mcelreath2020statistical, kruschke2013bayesian}. Specifying the priors for the parameters enables users to incorporate prior knowledge into the model and obtain stricter credible intervals, especially in the ties and order effects extensions discussed in the next section. 

This paper introduces the \texttt{bpcs} R package (Bayesian Paired Comparison in Stan) and the statistical models implemented in the package. This package aims to facilitate the use of Bayesian models for paired comparison data in research. The \texttt{bpcs} package provides a consistent interface for R users to work with these models and several functions for researchers to evaluate the posterior distribution of all parameters, to estimate the posterior distribution of any contest between items, and to obtain the posterior distribution of the ranks. The paper provides Bayesian reanalyses of three recent studies that used the frequentist Bradley-Terry model. These reanalyses are conducted with the Bayesian models of the \texttt{bpcs} package, and all the code used to fit the models, generate the figures, and the tables are available in the online appendix\footnote{\url{https://davidissamattos.github.io/bpcs-online-appendix/}}. 

\subsection{Related software}

This section includes relevant software for the Bayesian estimation of the Bradley-Terry and Thurstone models. \cite{johnson2013bayesian} provide a mathematical description, code and discussion about the implementation of Bayesian Thurstonian models for ranking data using the Gibbs Sampler JAGS.  However, it is still up the user of the software to process this data and generate relevant statistics, plots. Gibbs sampling is less efficient than the No-U-Turn Hamiltonian Monte Carlo (NUTS), used in the \texttt{bpcs} package. The NUTS is more efficient in terms of effective samples, in cases with higher autocorrelation and in hierarchical models \citep{nishio2019performance, carpenter2017stan, hoffman2014no}. 

\cite{caron2012efficient} proposes a specific Gibbs sampler for the generalized Bradley-Terry model. The proposed approach shows the Monte Carlo based samplers are efficient to estimate the parameters. However, their approach is limited to the generalized model and does extend to the extensions provide in the \texttt{bpcs} package.

The \texttt{sport} R package provides statistical models for sequential paired comparison data \citep{sport}, such as the Glicko and Glicko2  \citep{glickman2001dynamic} models and the Bayesian Bradley-Terry model. The Glicko and Glicko2 models are sequential and the results depend on the order in which the items are presented \citep{glickman2001dynamic, luckett2020estimates}. The package provides only the simple Bayesian Bradley-Terry model with a Bayesian Approximation Method. However, it does not allow to setup  priors or obtain the posterior distribution.

The \texttt{pcFactorStan} \citep{pritikin2020exploratory} R package implements a Bayesian Item Response Theory model for paired comparison. The items are measured with a Bayesian Bradley-Terry model and the worth values of the contestants are expanded with a latent factor score. While the \texttt{pcFactorStan} package can be used to create rank with the Bradley-Terry model, it focuses on answering factor analysis problems. Moreover, it does not provides the extensions for the generalized, hierarchical or model with ties.

The \texttt{thurstonianIRT} R package allows users to fit Item Response Theory models for forced-choice questionnaires. The package implements the models proposed by \cite{brown2011item, brown2016item}. The software utilizes different backend for model estimation (MPlus, laavan and Stan). While the package does not provide this functionality directly the Stan backend can provides estimation full posterior estimation of the Bayesian model.

The \texttt{PLMIX} implement finite mixture of Plackett-Luce models. In the case of partial rankings of two items (forced choice assessment) the Plackett-Luce model reduces to the Bradley-Terry model- \citep{turner2020modelling}. The package focuses on providing point estimates of the Bayesian estimation through Gibbs sampling. A part from handling ties, the package does not provide the additional extensions of the \texttt{bpcs} package.

\cite{corff2018bayesian} provides a custom non-parametric Gibbs sampling MCMC algorithm to approximate the posterior distribution of a Bayesian Bradley-Terry model in the random environment extension. While this extension is not considered in the \texttt{bpcs} package, the algorithm does not support the other use cases from the  \texttt{bpcs} or provide ready to use functions to support applied behavior research.

\cite{seymour2020bayesian} provide a Bayesian application of the Bradley-Terry model to spatial and geographical applications. The proposed extension introduces a multivariate normal prior distribution to model the spatial structure instead of linear regression  methods of generalized methods. The \texttt{bpcs} package utilizes a linear regression approach to include covariates. While it might not be suitable for the specific spatial application of \cite{seymour2020bayesian}, the linear model is the most common model and has been successfully used in applied research when including stimuli covariates \citep{dittrich1998modelling, fleischhaker2019modelling, giambona2020tourism}.

The \texttt{bpcs} package utilizes the No-U-Turn sampler implemented in Stan, provides an easy to use interface, a higher number of extensions, such as models with ties, generalized models, subject-specific predictors and hierarchical models,  functions to create tables, plots that facilitate interpretation of the models. Compared to the existing software packages, these features offer a higher flexibility, such as combining multiple extensions together, sampling efficiency using the No-U-Turn Hamiltonian Monte Carlo Sampler, and the easiness of use by providing a single consistent interface that hides the mathematical complexity of the models.

\section{Statistical models for paired-comparison}
\label{sec:models}

The mathematical models presented in this section have their origin in the Thurstone Law of Comparative Judgment \citep{thurstone1927law}. This law can assess the discriminal difference between two stimuli measured by a scale. \cite{thurstone1927law} proposes a series of cases in which assumptions are made to simplify the problem in terms of tractability \citep{tsukida2011analyze}.
In its general form (Case I), the estimation of the difference between two stimuli requires the estimation (or knowledge) of the dispersion of all stimuli and all of its correlations. The most known simplification (Case V) assumes that all options have equal dispersion and are uncorrelated. Due to its computational tractability, Case V has become more popular than the other less restrictive cases \citep{cattelan2012models, shah2015estimation}. Case V is also referred as the Thurstone-Mosteller model and often just as the Thurstonian model \citep{cattelan2012models, handley2001comparative, johnson2013bayesian}. 

The Bradley-Terry model \citep{bradley1952rank} provides a similar formulation to the Thurstone-Mosteller model, but assumes that the difference between two stimuli is a logistic random variable instead of a normally distributed variable \citep{cattelan2012models}.  In practice, both models yield nearly identical estimates and expected probabilities of one player beating the other \citep{handley2001comparative}. Unlike the Thurstone-Mosteller model, the Bradley-Terry model introduced an additional computational simplicity since the logit function has a closed-form expression.  Since its introduction in 1952, the Bradley-Terry model has been extended to encompass a different problems (order effects, random effects, ties, subject predictors, generalized models among others) in both the frequentist and Bayesian frameworks.

The \texttt{bpcs} implements the Bayesian extensions of the Bradley-Terry model. The choice for this family of models (instead of Thurstone Case I-V) is due to the wide use of Bradley-Terry models in behavior research, the large number of available extensions proposed in research, nearly identical estimates as the Thurstone Case V, and with better computational tractability.

This section provides an overview of the terminology, introduces the simplest case of the Bayesian Bradley-Terry model implemented in the package followed by a discussion of the different extensions.

\subsection{Terminology}

Different research areas work with different terminology and therefore it is worth having further clarifications:
\begin{itemize}
    \item Players or contestants are synonymous with the items being compared, i.e. the choices of some type of stimuli such as images, sounds, or objects. The \texttt{bpcs} package and this article utilize the term player.
    \item Subjects, participants, or judges are synonymous for the respondents of a questionnaire, or the subject that is selecting between the paired-comparison. Apart from the term multiple judgment sampling which refers to when a subject judges multiple items, the \texttt{bpcs} package and this article utilize the term subjects.
    \item A contest refers to a single comparison between two players made by a subject.
    \item Ties or draws refer to the case where a subject does not express a preference for a player in a contest. For example, in a questionnaire that asks subjects to select between two stimuli (the players) a tie could be an option such as ``I do not have a preference''. To avoid confusion withdrawing samples of a posterior distribution,  the \texttt{bpcs} package and this article utilize the term ties. 
\end{itemize}

The mathematical models utilize the following basic notation. Additional symbols and notations are presented with the extension that introduces them.
\begin{itemize}
    \item $\alpha_i > 0$: a latent variable that represents the ability of player $i$.
    \item $\lambda_i = \log(\alpha_i)$: the log of the ability of player $i$.
    \item $y_{i,j,n}$: the binary result of the contest $n$ between any two players $i$ and $j$. If  player $i$ wins, $y_{i,j,n}=1$. If player $j$ wins $y_{i,j,n}=0$. 
    \item tie$_{i,j,n}$: a binary variable representing if the result of a single contest (at position $n$) between two players ($i$ and $j$) was a tie. It assumes tie$_n=1$ if it was a tie and tie$_k=0$ if it was not a tie.
    \item $N$: the number of players.
    \item $\sigma^2_{\lambda}$: the variance of the normal prior distribution for the variable $\lambda$. Analogously goes for the prior distributions for the other parameters $\gamma$, $\nu$, $\beta$ and $U$.
\end{itemize}

To simplify the notation of the presented models below the index $n$ is omitted.

\subsection{The Bradley-Terry model}
The Bradley-Terry model \citep{bradley1952rank} presents a way to calculate the probability of one player beating the other player in a contest. This probability is represented by:

\begin{equation}
     \pp [i \text{ beats } j] = \dfrac{\alpha_i}{\alpha_i+\alpha_j} 
\end{equation}

This model is commonly parameterized by the log of the ability variable:

\begin{equation}
     \pp [i \text{ beats } j] = \dfrac{\exp{\lambda_i}}{\exp{\lambda_i} +\exp{\lambda_j}} 
\end{equation}

This transformation have the benefits of allowing the estimation of a parameter $\lambda \in (-\infty, \infty)$ and simplifying the estimation of the parameters for the frequentist setting with a generalized linear models with the logit function \citep{cattelan2012models}:

\begin{equation}
     \text{logit}(\pp [i \text{ beats } j]) = \lambda_i - \lambda_j
\end{equation}

However, the parameters $\lambda$ are not uniquely identified and require another constraint, commonly:

\begin{equation*}
        \sum ^{N} _{i=1} \lambda_i = 0 \text{ or } \sum ^{N} _{i=1} \lambda_i = 1
\end{equation*}

The first work to propose a Bayesian formulation of the Bradley-Terry model is attributed to Davidson and Solomon \citep{davidson1973bayesian}. They proposed a version of the Bradley-Terry model utilizing a conjugated family of priors and estimators to calculate the posterior distribution of the log abilities of the parameters and the rank of the players. \cite{leonard1977alternative} discusses the issue of the flexibility an interpretation of the conjugated priors proposed by Davidson and Solomon in the presence of additional explanatory variables and other extensions. 

\cite{leonard1977alternative} suggests moving away from the convention of using conjugated priors and utilizing normal prior distributions for the parameters. The usage of non-conjugated prior distributions has many advantages, including adaptability to extensions, being able to reason and fully specify the prior parameters, ability to extend to hierarchical models, and to specify other prior distribution families. 

The two main disadvantages of the approach proposed by Leonard (1977) is the use of approximation methods for the posterior distribution and the computational time. However, the advances in Bayesian computational packages and Markov Chain Monte Carlo (MCMC) samplers significantly minimize such disadvantages. The \texttt{bpcs} package utilizes normal priors for the $\lambda$ parameters and models the outcome variable $y_{i,j}$ of a single contest between players $i$ and $j$ with a Bernoulli distribution, based on the probability of winning for $\pp [i \text{ beats } j]$.

Therefore, the simple Bayesian Bradley-Terry model can be represented as:

Likelihood:
\begin{align}
    \pp [i \text{ beats } j] &= \dfrac{ \exp{\lambda_i} }{  \exp{\lambda_i} +  \exp{\lambda_j}} \\
    y_{i,j} &\sim \text{Bernoulli}(\pp [i \text{ beats } j])
\end{align}

Priors:
\begin{align}
    \lambda_i &\sim \nn(0, \sigma^2_{\lambda})
\end{align}

The mean of the prior distribution changes the location of the parameters without impacting the relative probability of one player beating the other. Since it does not impact the estimation of the $\lambda_i$, its value is set to zero.
The standard deviation in the prior distribution represents the space where the model should look for the relative differences in the probabilities. Smaller values for the standard deviations indicate that the relative preferences are close and have many overlap, higher standard deviation indicates that the sampler can look for solutions that are very far apart (probabilities closer to zero or 1). Choosing high standard deviations for the prior can increase the time to find a solution and possibly divergence between the chains in the sampler. Very low standard deviations are informative priors and imply that the strength parameters are very close to each other.  Both the mean and the standard deviation of the prior acts as soft constraint, making the model identifiable \citep{pritikin2020exploratory, stan2016stan}.

The parameters $\lambda_i$ can be used to rank the different players. However, in the Bayesian framework, a single measure is not obtained but rather a posterior distribution of the parameters $\lambda_i$. By sampling from the posterior distribution of the log-abilities of the players, it is possible to create a posterior distribution of the ranks of the players, which helps to evaluate the uncertainty in the ranking system.

\subsection{Davidson model}
The first extension to be added to the Bradley-Terry model is the ability of handling ties in the contest between two players. This approach was proposed by \cite{davidson1970extending}, which adds an additional parameter $\nu$ and computes two probabilities: the probability of $i$ beating $j$ given that it was not a tie $\pp [i \text{ beats } j \vert \text{not tie}]$  and the probability of the result being a tie $\pp [i \text{ ties } j]$. A Bayesian formulation of the model is represented below:

Likelihood:
\begin{align}
    \pp [i \text{ beats } j \vert \text{not tie}] &= \dfrac{ \exp{\lambda_i} }{  \exp{\lambda_i} +  \exp{\lambda_j{ +  \exp{(\nu + \frac{\lambda_i + \lambda_j}{2}) }}}} \\
    \pp [i \text{ ties } j] &= \dfrac{ \exp{(\nu + \frac{\lambda_i + \lambda_j}{2}) } }{  \exp{\lambda_i} +  \exp{\lambda_j} + \exp{(\nu + \frac{\lambda_i + \lambda_j}{2} )}} \\
    y_i &\sim \text{Bernoulli}(\pp [i \text{ beats } j \vert \text{not tie}]) \\
    \text{tie}_{i,j} &\sim \text{Bernoulli}( \pp [i \text{ ties } j] )
\end{align}    
   
Priors: 
\begin{align}
      \lambda_i &\sim \nn(0, \sigma^2_{\lambda})\\
    \nu &\sim \nn(0, \sigma^2_{\nu})  
\end{align}

The $\nu$ parameter in (the log scale) balances the probability of ties against the probability of not having ties. If $\nu \to -\infty$ than $\pp [i \text{ ties } j] \to 0$ regardless of the players abilities, which is equivalent to the Bradley-Terry model. If  $\nu \to +\infty$ than $\pp [i \text{ ties } j] \to 1$ regardless of the players abilities. If $\nu = 0$ than $\pp [i \text{ ties } j] = \frac{ \exp{\frac{\lambda_i + \lambda_j}{2} } }{  \exp{\lambda_i} +  \exp{\lambda_j} + \exp{\frac{\lambda_i + \lambda_j}{2} }}$, which means that the probability of ties depends only on the players abilities. In this last case, if the players $i$ and $j$ have equal abilities, they have equal chance of having a tie, $i$ winning or $j$ winning.

For the Bayesian version of the Davidson model, the choice of prior in the $\nu$ parameter refers to the prior belief on how ties are affected by the relative difference in players' abilities. If the intended goal is also to investigate if the probability of wins and ties depend only on the abilities, the mean parameter of the normal prior distribution for $\nu$ is set to zero (as in the presented models).

Although the subsequent models are presented only in the Bradley-Terry variation, they can be easily extended and implemented as a Davidson model to handle ties. 

\subsection{Models with order effect}
In paired comparison problems, a problem that can arise is that the order in which two players are presented can lead to a bias in the choice of the comparison. For example, when two items are presented to be chosen, subjects might have a preference for items placed on the left side. Another example that arises from sports competitions is the presence of home-field advantage, athletes (the players in paired-comparison) competing in their home field can have an advantage compared to the visitor player. 

The order effect can be modeled as either additive or multiplicative. \cite{davidson1977extending} discuss some of the advantages of the multiplicative model compared to the additive. One important advantage of the Bayesian version is that the value of the order effect parameter does not depend on the abilities parameters. Therefore, setting the prior distribution of the order-effect parameter is independent of the soft-constraints applied to the log-abilities parameter. Additionally, the multiplicative model has easily introduced to both the Bradley-Terry and the Davidson model when estimating the parameters in the log scale.

To compensate for the order effect using a multiplicative model, \cite{davidson1977extending} introduced an additional parameter $\gamma$. This multiplicative parameter becomes in the log-scale an additive term as shown below. The Bayesian Bradley-Terry model with order effect can be represented as:

Likelihood:
\begin{align}
    \pp [i \text{ beats } j] &= \dfrac{ \exp{\lambda_i} }{  \exp{\lambda_i} +  \exp{(\lambda_j + \gamma})} \\
    y_i &\sim \text{Bernoulli}(\pp [i \text{ beats } j])
\end{align}

Priors: 
\begin{align}
    \lambda_i &\sim \nn(0, \sigma^2_{\lambda})\\
    \gamma &\sim \nn(0, \sigma^2_{\gamma})
\end{align}

The $\gamma$ parameter in (the log scale) reflects the impact of the order effect. If $\gamma \to -\infty$ than $ \pp [i \text{ beats } j] \to 1$ which means that regardless of the players' abilities, the player $i$ will have an order effect advantage and will always win the contest. Analogously  if  $\gamma \to +\infty$ than $ \pp [i \text{ beats } j] \to 0$ and player $i$  will have an order effect disadvantage and will always lose the contest. If $\gamma = 0$ then player $i$ will neither have an advantage or disadvantage with the order effect, and the probability of wins or ties depends only on the players' abilities and the tie parameter $\nu$. The choice of prior in the $\gamma$ parameter refers to our prior belief on the location and spread of the value of order effect. If the intended goal is also to investigate if there is an order effect or not, the mean parameter of the normal prior distribution for $\gamma$ is set to zero (as in the presented models).

\subsection{Generalized models}
Many research problems require the investigation of the effect of players properties in the probability of winning a contest. The extension proposed by \cite{springall1973response} is analogue to the multiple regression case. This extension proposes the use of $K$ predictors that are characteristic of the players (and not of the subjects which is discussed later). $X_{i,k}$ is the $k$ predictor value of player $i$ and $\beta_k$ is the coefficient of the predictor that we are estimating. Note that for these generalized models the intercept is not identifiable and therefore not included \citep{springall1973response, stern2011moderated}. The Bayesian generalized Bradley-Terry can be represented as:

Likelihood:
\begin{align}
    \pp [i \text{ beats } j] &= \dfrac{ \exp{\lambda_i} }{  \exp{\lambda_i} +  \exp{\lambda_j}} \\
    \lambda_i &= \sum_k^N X_{i,k}\beta_k \\
     y_i &\sim \text{Bernoulli}(\pp [i \text{ beats } j])
\end{align}

Priors:
\begin{align}
    \beta_k &\sim \nn(0, \sigma^2_{\beta})
\end{align}

The generalized version of the Bradley-Terry model estimates the parameters $\beta_k$. The parameter $\lambda_i$ is then estimated by the linear model $\lambda_i = \sum_k^N X_{i,k}\beta_k$. The choice of prior in the $\beta$ parameter refers to the prior belief on the value of the coefficient of each predictor. The presented generalized models utilize the same prior for all $\beta$ coefficients. Therefore it requires that the values for every $k$ in the $X_{i,k}$ be on the same range, otherwise, the model would have a strong informative prior belief in some coefficients and a weakly informative prior belief in other coefficients. If the predictors' input values $X_{i,k}$ are normalized for every $k$ the larger the coefficient $\beta_k$, the higher the influence of that predictor in the probability of winning.

\subsection{Introducing random-effects to model-dependent data}
It is common in the research context to have the same subject to make multiple comparisons, the multiple judgment sampling problem \citep{cattelan2012models}. A more realistic analysis of the Bradley-Terry model would assume that the comparisons made by the same person are dependent. One approach to address the multiple judgment sampling problems is through the usage of mixed-effects or hierarchical paired comparison models \citep{bockenholt2001hierarchical}. 

\cite{bockenholt2001hierarchical} decomposed the paired comparison model into fixed and a random effect components. The random effects component estimates the subject variation (given $S$ subjects) in each item, while the fixed effect component estimates the average log ability of the player. The random effects term is represented by $U_{i,s}$, where $i$ refers to the player being judged and $s$ to the subject. The Bayesian Bradley-Terry model with random effects can be represented as:

Likelihood:
\begin{align}
    \pp [i \text{ beats } j] &= \dfrac{ \exp{\lambda_{i,s}} }{  \exp{\lambda_{i,s}} +  \exp{\lambda_{j,s}}} \\
    \lambda_{i,s} &= \lambda_i + U_{i,s}\\
    y_{i,j} &\sim \text{Bernoulli}(\pp [i \text{ beats } j])\\
\end{align}

Priors:
\begin{align}
    \lambda_i &\sim \nn(0, \sigma^2_{\lambda}) \\
    U_{i,s} &\sim \nn(0,U^2_{\text{std}})\\
    U_{\text{std}} &\sim \text{Half-}\nn(0, \sigma^2_{U})
\end{align}

This model aims at estimating the parameter $U_{\text{std}}$ that represents the standard deviation in the random effects and the difference between subjects. In the Bayesian context, with the Stan probabilistic programming language, it is also possible to estimate the parameters $U_{i,s}$ (a total of $SN$ parameters). The choice of prior for the $U_{\text{std}}$ represents the prior belief in the difference in judgment between the subjects. It can be set to be a weakly-informative prior (with a large value for $\sigma^2_{U}$). The prior distribution for $U_{\text{std}}$ is a half-normal distribution, i.e., normal distribution where only values above zero are valid.

\subsection{Subject-specific predictors}
The last extension presented in this article, is the inclusion of subject-specific covariates. In many behavior research problems is desired to evaluate how characteristics of the subject influences the choice in a contest. This extension was originally proposed by \cite{bockenholt2001hierarchical} models  subject-specific covariates for each player utilizing a linear regression. This model utilizes the following notation: $K$ is the number of subject-specific predictors, $x_{i,k,s}$ representing the observed covariate $k$ of subject $s$ for player $i$ and the coefficient for the covariate $k$ of player $i$  represented by $S_{i,k}$. The model can be represented as:

Likelihood:
\begin{align}
    \pp [i \text{ beats } j] &= \dfrac{ \exp{\lambda_{i,s}} }{  \exp{\lambda_{i,s}} +  \exp{\lambda_{j,s}}} \\
    \lambda_{i,s} &= \lambda_i + \sum ^K _{k=1} x_{i,k,s}S_{i,k}\\
    y_{i,j} &\sim \text{Bernoulli}(\pp [i \text{ beats } j])
\end{align}

Priors:
\begin{align}    
    \lambda_i &\sim \nn(0, \sigma^2_{\lambda}) \\
    S_{i,k} &\sim \nn(0,\sigma^2_{S})
\end{align}

These models estimate the baseline ability of the players, $\lambda_i$, and the subject-specific coefficients of the covariates $S_{i,k}$. These coefficients represent how a change in the subject covariate for each player will impact the probability of selecting player $i$ over player $j$. The model estimates one coefficient for every covariate of every player, resulting in a total of $K\cdot N$ estimated coefficients. It is worth noting that covariates coefficients are specific to each player and therefore can have a different impact depending on how it influences the player log ability. These coefficients can be used to investigate systematic differences in how each player is evaluated by the subjects. 

It is worth noting, that again the absolute values of these coefficients do not have a direct interpretation of the effect it adds to the probability of one player beating another. This effect depends on the relative impact of the covariate in the two players and it is better assessed through the actual probability of selecting one player over the other. In the Bayesian context, this can be assessed through the posterior distribution of the probabilities and the absolute effect can be measured.

These models assume that the covariates $x_{i,k,s}$ have a similar range of values and that are centered since they utilize the same normal priors with zero mean and constant standard deviation. In practice, this means that the values of the coefficients are more easily estimated by the MCMC sampler if all values of $x_{i,k,s}$ is normalized by each covariate. This model accepts both categorical predictors as well as continuous predictors. Categorical predictors can be added utilizing dummy-coding.

\subsection{Remarks}
Although not presented here, all the discussed extensions can be incorporated in a single model mathematical model, since they are all linearly added in the exponential terms. The \texttt{bpcs} package can handle from both the simple Bradley-Terry and the Davidson model to any combination of these extensions to these models.

Even in more complex, models the interpretation of the extension parameters remains the same as presented here. However, it is worth reinforcing that these parameters should always be analyzed in the context of the effect sizes, i.e., the actual probabilities of one player beating the other given the changes in the other parameters.

\section{The bpcs package}
\label{sec:bpcs}

This section presents a short overview of the underlying implementation of the \texttt{bpcs} package and its main functionalities. The \texttt{bpcs} R package implements the Bayesian version of the Bradley-Terry model and its extensions, as discussed in the statistical models' section. The models are coded in the Stan language and utilize the No-U-Turn (NUTS) Hamiltonian Monte Carlo sampler \citep{hoffman2014no}, which provides several advantages over the Gibbs sampler \citep{nishio2019performance, carpenter2017stan, hoffman2014no}. The latest version of the package and installation instructions can be found in the package repository\footnote{\url{https://github.com/davidissamattos/bpcs}}.

\subsection{Basic usage}
To exemplify the basic usage of the \texttt{bpcs} package,  \cite{luckett2020estimates} work is used as an example.The authors investigate the relative acceptability of food and beverage choices using paired preferences. One of the examples discussed is the acceptability of five brands of four cheese frozen pizzas. The full code for this presentation of the package and the reanalyses are available in the online appendix.

The main function of the \texttt{bpcs} package is the \texttt{bpc} function. The \texttt{bpc} function takes as input arguments: a data frame, two string columns with the names of contestants, a string with the result of the contestant (0 for player0, 1 for player1, or 2 for ties), and the model type. The model type is specified with a string. Two basic models are available, the '\textit{bt}' model for the Bradley-Terry model \citep{bradley1952rank}, and the '\textit{davidson}' model for the Davidson model to handle ties \citep{davidson1970extending}. Extensions for each of these base models can be added using a dash separator and the extension, for example, '\textit{bt-ordereffect}' specifies the Bradley-Terry model with order effect; '\textit{davidson-generalized-U}' specifies the generalized Davidson model including random effects. All presented extensions in the statistical models' section can be added to both base models, including more than one extension at the same time.

Other options, such as the method for handling ties, calculating the results from the scores of each player, column for clusters, specification of the priors, number of iterations to sample, among others, are described in the documentation\footnote{\url{https://davidissamattos.github.io/bpcs/}}. The call for the \texttt{bpc} function is shown in the listing \ref{lst:bpc}:

\begin{lstlisting}[caption={The bpc function}, label={lst:bpc}]
m <- bpc(data = dpizza,
         player0 = 'Prod0',
         player1 = 'Prod1',
         result_column = 'y',
         solve_ties = 'none',
         model_type = 'bt',
         iter=3000)
\end{lstlisting}

The package also implements the S3 functions \texttt{print}, \texttt{summary}. \texttt{plot} and \texttt{predict}. The \texttt{print} function displays the parameters table with the High Posterior Density Intervals \citep{kruschke2018bayesian, mcelreath2020statistical}. \texttt{summary} function prints the parameters table, a table with a posterior probability of winning for all combination of players and a posterior rank of the players including the median rank, mean rank, and the standard deviation of the rank. The \texttt{plot} function provides a caterpillar plot of the model parameters with the correspondent HPD or credible intervals. The \texttt{predict} function provides a posterior distribution of predictive results of any match between the players of the fitted model. Below is the result of the \texttt{summary} function for the model.

\begin{lstlisting}[caption={Output of the summary function}, label={lst:summary}]
Estimated baseline parameters with 95% HPD intervals:

Table: Parameters estimates

Parameter              Mean   Median   HPD_lower   HPD_higher
-------------------  ------  -------  ----------  -----------
lambda[Tombstone]     -0.14    -0.11       -2.77         2.57
lambda[DiGiorno]       0.33     0.34       -2.30         3.03
lambda[Freschetta]     0.22     0.25       -2.42         2.97
lambda[Red Barron]     0.24     0.26       -2.41         2.95
lambda[aKroger]       -0.35    -0.32       -2.97         2.39
NOTES:
* A higher lambda indicates a higher team ability

Posterior probabilities:
These probabilities are calculated from the predictive posterior distribution
for all player combinations


Table: Estimated posterior probabilites

i            j             i_beats_j   j_beats_i
-----------  -----------  ----------  ----------
aKroger      DiGiorno           0.39        0.61
aKroger      Freschetta         0.35        0.65
aKroger      Red Barron         0.31        0.69
aKroger      Tombstone          0.47        0.53
DiGiorno     Freschetta         0.54        0.46
DiGiorno     Red Barron         0.48        0.52
DiGiorno     Tombstone          0.64        0.36
Freschetta   Red Barron         0.44        0.56
Freschetta   Tombstone          0.67        0.33
Red Barron   Tombstone          0.60        0.40

Rank of the players' abilities:
The rank is based on the posterior rank distribution of the lambda parameter

Table: Estimated posterior ranks

Parameter     MedianRank   MeanRank   StdRank
-----------  -----------  ---------  --------
DiGiorno               1       1.65      0.78
Freschetta             2       2.24      0.85
Red Barron             2       2.20      0.84
Tombstone              4       4.07      0.52
aKroger                5       4.84      0.37
\end{lstlisting}

The package also provides helper functions to create plots and, to generate formatted tables (such as the ones from the \texttt{summary} function) for Latex, HTML, and the Pandoc\footnote{\url{https://pandoc.org/}} format (which in turn can be used to generate Microsoft Word tables). These functions are:

\begin{itemize}
    \item \texttt{get\_parameters\_table} This function generates a table of the parameters with summary statistics of the posterior distribution. Two measures of uncertainty are available, the equal-tailed intervals and Highest Posterior Density (HPD) intervals \citep{kruschke2018bayesian, mcelreath2020statistical}. The equal-tailed intervals divide the posterior distribution into two parts with the same probability mass, i.e. both tails have the same probability of being selected.  The HPD interval corresponds to the narrowest interval that contains the mode for a unimodal distribution. In the case of a symmetrical unimodal distribution (such as the normal distribution), both intervals are equivalent. However, in the case of a non-symmetrical distribution, these intervals will be different and the HPD interval will be shorter.
    \item \texttt{get\_probabilities\_table}. This function generates a table of the probabilities of one player being chosen against another player. These probabilities are calculated by sampling the predictive posterior distribution of the results.
    \item \texttt{get\_rank\_of\_players\_table}. This function calculates the rank of each player based on the posterior distribution of the log-abilities of the players (the $\lambda$). By assessing the posterior distribution of the rank and looking at the standard deviation, it is possible to assess the uncertainty on the rank estimates. Estimating the uncertainty in the rank values is not available in any of the frequentist packages.
    \item \texttt{plot}. This function creates a caterpillar type of plot of the log-ability parameters of the players with the uncertainty intervals. This function returns a \texttt{ggplot2} \citep{ggplot2} object which can be easily customized by the user.
\end{itemize}

If the user has a higher need to customize the tables, the user can either provide further customization with additional packages such as the \texttt{kableExtra} package \citep{kableExtra}, or the utilize the function \texttt{get\_parameters\_df}, \texttt{get\_probabilities\_df} or \texttt{get\_rank\_of\_players\_df} to obtain a data frame that contains only the data of the table. The online appendix utilizes these approaches to create more complex tables for the reanalyses.

\subsection{Model validity}

After the call of the \texttt{bpc} function, the \texttt{bpcs} package runs the No-U-Turn Hamiltonian Monte Carlo sampler \citep{hoffman2014no} from Stan to estimate the posterior distribution of the parameters of the model. Before interpretation of the results, the user should check if the model has converged and or if there were problems in the convergence. If the model has not converged properly the posterior distribution should not be interpreted. The basic checks are:

\begin{itemize}
    \item Properly mixed chains. When sampling, it is common to use multiple chains. The chains should converge to the same value for every parameter and should not show any visible pattern \citep{mcelreath2020statistical}. A good convergence has a stationary caterpillar format. This can be checked using traceplots. Chains that have not converged in the presented paired comparison models are usually due to very large variance on the priors (which can lead to unidentifiable models since the soft-constraint is not sufficient). 
    \item Gelman-Rubin convergence coefficient (split $\hat{R}$). This coefficient is another measure of convergence \citep{gelman1992inference}. A value close to 1 indicates indicate that the chains have converged to the same values. In practical terms, values of $\hat{R}<1.01$ are required to indicate a good convergence \cite{mcelreath2020statistical, vehtari2021rank}.
    \item Number of diverging iterations. Diverging iterations indicate that the sampler has not completely explored the solution space for the posterior. If there are diverging iterations the results can be biased \citep{betancourt2017conceptual}. A common solution is to increase the number of iterations for the warmup and the target acceptance probability parameter.
    \item Number of effective samples. This diagnoses the precision of the sampler estimation \citep{zitzmann2019going}. The number of effective samples of the posterior indicates the number of independent samples. As a rule of thumb, 200 effective samples of the posterior are enough to estimate the mean of a parameter but more is required for estimating extreme quantiles \citep{zitzmann2019going, mcelreath2020statistical}.
\end{itemize}

While these are the basic checks for any Monte Carlo sampler, there are two additional diagnostics specific to the Hamiltonian Monte Carlo (HMC) sampler used by the \texttt{bpcs} package.

\begin{itemize}
    \item Maximum treedepth limits: the HMC imposes a limit in the depth of the trees that it evaluates at each iteration. If this limit is hit, it indicates that the sampler terminated to avoid long execution times. While it does not present a validity concern, the maximum treedepth represents an efficiency concern in terms of execution time. In the absence of other problems, increasing the treedepth may correct the problem \citep{stan2016stan}. 
    \item Low potential energy: a low kinetic energy calculated by the Estimated Bayesian Fraction of Missing Information (E-BFMI) indicates that the the chains have not explored the posterior distribution efficiently. If this occurs a common solution is to run the model for more iterations \citep{stan2016stan, betancourt2017conceptual}.
\end{itemize}

The \texttt{bpcs} package offers basic convergence checks with the function \texttt{check\_convergence\_diagnostics}, as shown below. Figure \ref{fig:traceplots} shows the traceplots for the pizza model.

\begin{lstlisting}[caption={Output of the check\_convergence\_diagnostics function.}, label={lst:checks}]
> check_convergence_diagnostics(m)

Checking sampler transitions treedepth.
Treedepth satisfactory for all transitions.

Checking sampler transitions for divergences.
No divergent transitions found.

Checking E-BFMI - sampler transitions HMC potential energy.
E-BFMI satisfactory for all transitions.

Effective sample size satisfactory.

Split R-hat values satisfactory all parameters.

Processing complete, no problems detected.
\end{lstlisting}

\begin{figure}
\centering
\includegraphics[width=\textwidth]{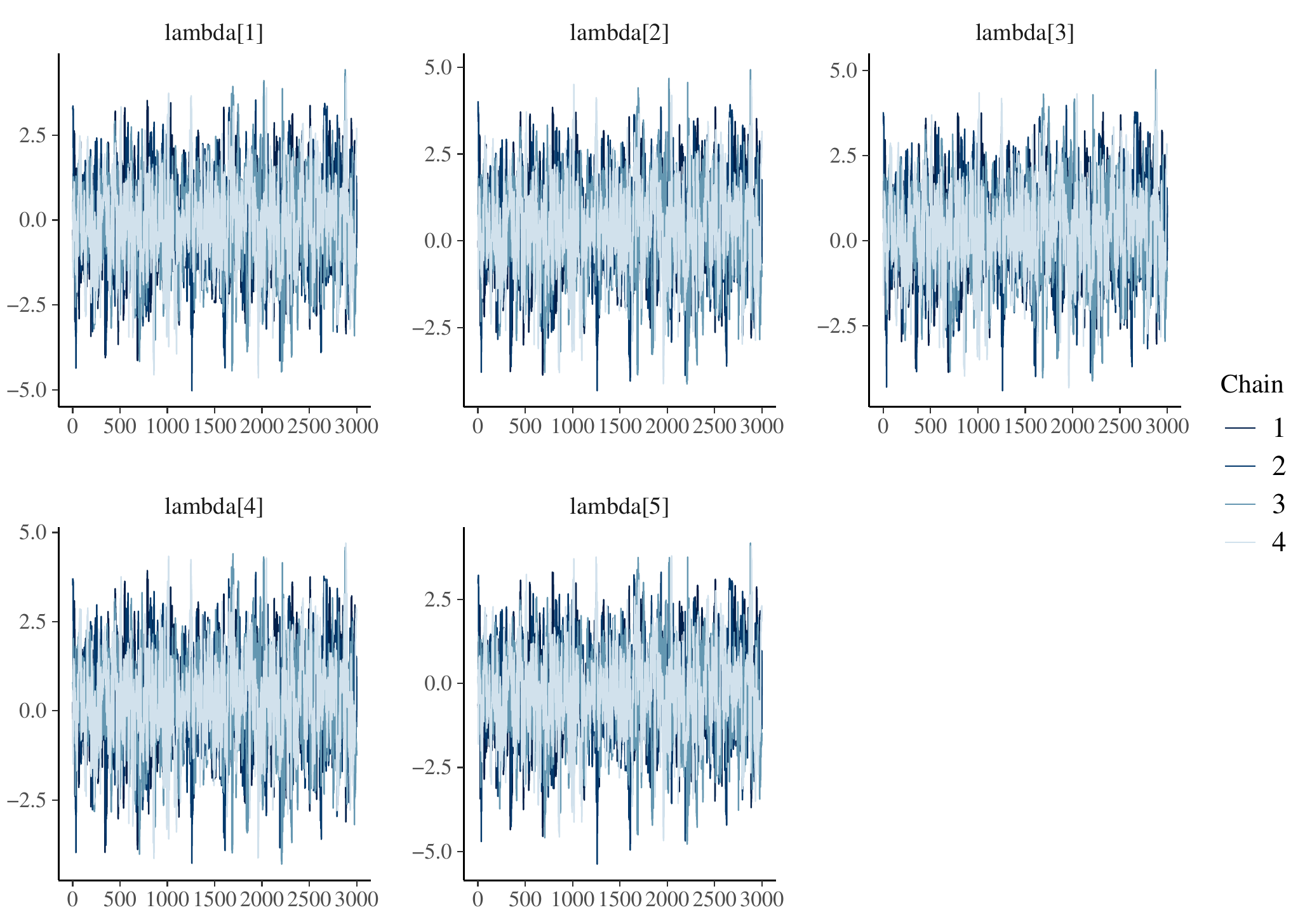}
\caption{\label{fig:traceplots} Example of traceplots for the parameters of the pizza model. Note that the traceplots do not contain any apparent pattern (are stationary) and all chains are overlapping in a caterpillar format.}
\end{figure}

These checks among other plots can also be verified with the \texttt{shinystan} package \citep{shinystan}. This package provides a web-based graphical user interface that implements convergence and posterior checks. The interface can be launched directly from the \texttt{bpcs} package with the function \texttt{launch\_shinystan}.

\subsection{Model comparison}

Bayesian statistics reinforces generating several valid models that can explain the obtained data and comparing them \citep{mcelreath2020statistical}. One approach to compare these models is with the use of an information criterion, such as the Watanabe-Akaike Information Criterion (WAIC) \citep{gelman2014understanding} or the Leave-One-Out Cross-Validation method (LOO-CV) \citep{loo}. 

The \texttt{bpcs} package provides both of these estimates with the functions \texttt{get\_waic} and \texttt{get\_loo} as shown below. 

\begin{lstlisting}[caption={WAIC and LOO-CV in the bpcs package.}, label={lst:waic}]
> get_waic(m)

Computed from 12000 by 380 log-likelihood matrix

          Estimate  SE
elpd_waic   -259.8 3.9
p_waic         4.0 0.1
waic         519.6 7.8
> get_loo(m)

Computed from 12000 by 380 log-likelihood matrix

         Estimate  SE
elpd_loo   -259.8 3.9
p_loo         4.0 0.1
looic       519.6 7.8
------
Monte Carlo SE of elpd_loo is 0.0.

All Pareto k estimates are good (k < 0.5).
See help('pareto-k-diagnostic') for details.
\end{lstlisting}

Note that the information criteria Akaike Information Criterion (AIC) and the Bayesian Information Criterion (BIC) should not be used since they assume models with flat priors and maximum a posteriori estimates \citep{mcelreath2020statistical}. These assumptions are not valid in the models implemented in the \texttt{bpcs} package and therefore the package does not provide these estimates.

\subsection{Limitations of the bpcs package}
The main limitation of these Bayesian paired comparison methods is the computational costs. While for most research problems a standard laptop should be able to create a posterior distribution of the parameters of the model in a few minutes, there are specific problems that will require more computational power, or even a transition from the Bayesian models towards frequentist models, if the posterior distribution is not used. For example, \cite{zhang2019establishing} utilizes four Bradley-Terry models to rank a total of 7,035 images with an order of 120,000 data points for each model. While it is still possible to perform Bayesian inference in this problem, fitting the model might take several hours. However, for many practical research problems, Bayesian Bradley-Terry models might take only minutes.


\section{Reanalyses}
\label{sec:reanalysis}

This section provides Bayesian reanalyses of three studies conducted with a frequentist implementation of the Bradley-Terry model. The commented code to generate the figures and tables of these reanalyses are available in the appendix. These reanalyses provide information regarding what was presented in the original papers, followed by discussions of alternative models. However, the reanalyses do not cover all possible models available in the \texttt{bpcs} package, such as the models with ties and generalized models. Examples of these models in other areas are provided in the package documentation\footnote{\url{https://davidissamattos.github.io/bpcs/}}.

\subsection{Study I: Visual Perception of Moisture Is a Pathogen Detection Mechanism of the Behavioral Immune System}

This reanalysis is based on the study ``Visual Perception of Moisture Is a Pathogen Detection Mechanism of the Behavioral Immune System'' \citep{iwasa2020visual}. In this study, the authors utilized paired comparisons to rate the perceived moisture content based on the visual perception for high luminance areas in images. The participants were asked to select the image that had the highest moisture content.  The paired images were presented twice for each participant first left to right and then right to left, to control for the influence of the presentation position. The image stimuli are presented in the supplementary material of the original article. 

This reanalysis replicates the results of the study with a Bayesian Bradley-Terry model and then investigates the presence and magnitude of the order-effect. The two models are compared utilizing the Watanabe-Akaike information criterion (WAIC). 

\begin{figure}
\centering
\includegraphics[width=\textwidth]{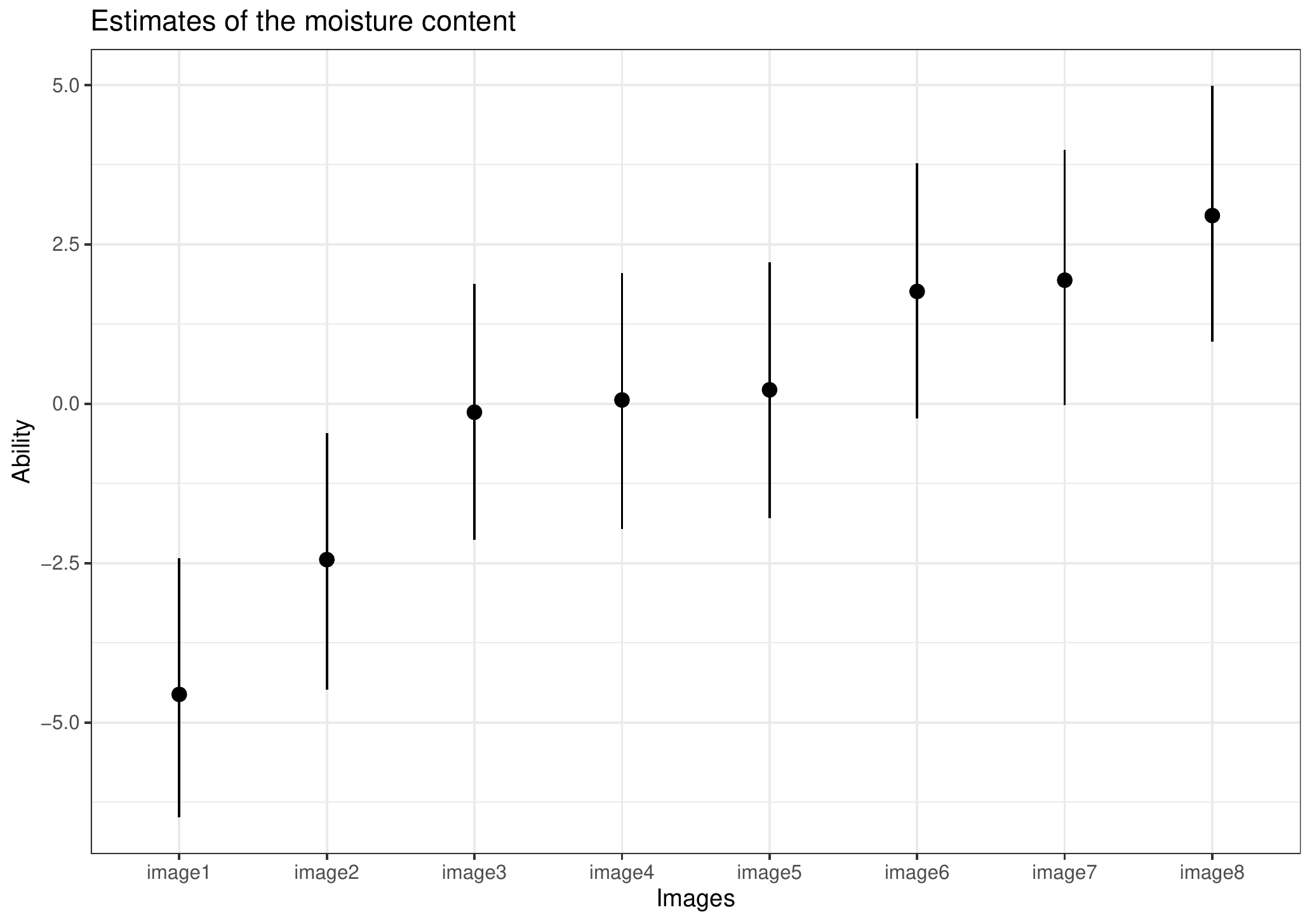}
\caption{\label{fig:r1-plot} Worth values of the images, in terms of moisture content, and their respective 95\% HPD interval in the simple Bradley-Terry model.}
\end{figure}

For the simple Bradley-Terry model, Table \ref{tab:r1-m1-parameters} shows the worth value of each parameter that indicates the moisture content and the number of effective samples. Figure \ref{fig:r1-plot} shows the parameter plot with the 95\% HPD intervals. The WAIC of the model is equal to 8132.2.

The second model adds an order-effect term to the model. The posterior estimation of the $\gamma$ parameter is close to zero (with mean -0.001, lower HPD -0.054, and upper HPD 0.056), which indicates that there is no presence of order effect. The WAIC of the model is equal to 8134.1. The WAIC of the order effect is higher than the WAIC of the simple Bradley-Terry model, which indicates that the additional parameter did not increase the predictive values of the model. This indicates that, for this study, the strategy to show both images twice with change in the presentation order was effective to control for order effects. For the remaining analysis, the selected model is the Bradley-Terry model without order effect.

The priors were chosen to be normal distributions centered around 0 and with variance of 3.0. This variance allows probabilities of $i$ beating $j$ up to (given that each player can be up three standard deviations from the mean in each extreme) 0.99997. While this prior still regularizes and makes the model identifiable it is considered weakly informative. The prior for the order effect $\gamma$ parameter was set to mean zero and variance of 1.0. This prior indicates that the order effect can be for both the right or left images.

\begin{table}[ht]
\caption{\label{tab:r1-m1-parameters}Parameters estimates for the simple Bradley-Terry model}
\centering
\begin{tabular}[t]{lrrrrr}
\toprule
Parameter & Mean & Median & HPD lower & HPD upper & N. Eff. Samples\\
\midrule
image1 & -4.577 & -4.554 & -6.533 & -2.489 & 598\\
image2 & -2.465 & -2.436 & -4.503 & -0.461 & 596\\
image3 & -0.154 & -0.120 & -2.221 & 1.830 & 596\\
image4 & 0.038 & 0.069 & -2.065 & 1.977 & 596\\
image5 & 0.197 & 0.227 & -1.874 & 2.175 & 595\\
image6 & 1.742 & 1.770 & -0.349 & 3.668 & 594\\
image7 & 1.917 & 1.947 & -0.188 & 3.842 & 594\\
image8 & 2.930 & 2.967 & 0.879 & 4.920 & 597\\
\bottomrule
\end{tabular}
\end{table}

Considering the estimates of the first model in Figure \ref{fig:r1-plot}, it is possible to identify a large overlap in the interval between the latent worth value of some image groups (image3, image4, image5 and also image6, image7). However, to properly rank and differentiate them, it is necessary to generate a posterior distribution of the rank of these images. Table \ref{tab:r1-m1-rank} ranks the images based on the posterior distribution of the ranks in terms of moisture content. From this table, it is possible to see that despite the large overlap in the HPDI of the worth values, the images differentiate themselves in distinct ranks with low variation in the ranks.

\begin{table}[ht]
\caption{\label{tab:r1-m1-rank} Rank of the images based on moisture content.}
\centering
\begin{tabular}[t]{lrrr}
\toprule
Parameter & Median Rank & Mean Rank & Std.Rank\\
\midrule
image8 & 1 & 1.00 & 0.00\\
image7 & 2 & 2.00 & 0.05\\
image6 & 3 & 3.00 & 0.05\\
image5 & 4 & 4.01 & 0.08\\
image4 & 5 & 5.00 & 0.08\\
image3 & 6 & 6.00 & 0.03\\
image2 & 7 & 7.00 & 0.00\\
image1 & 8 & 8.00 & 0.00\\
\bottomrule
\end{tabular}
\end{table}

\subsection{Study II:  Using a Touchscreen Paradigm to Evaluate Food Preferences and Response to Novel Photographic Stimuli of Food in Three Primate Species}

This reanalysis is from the article ``Using a Touchscreen Paradigm to Evaluate Food Preferences and Response to Novel Photographic Stimuli of Food in Three Primate Species (Gorilla gorilla gorilla, Pan troglodytes, and Macaca fuscata)'' \citep{huskisson2020using}, an extension of the initial study with a single gorilla  \citep{hopper2019assessment}. In this study, the authors tested a protocol of pairwise forced choice with six stimuli of food (four familiar and two novel stimuli) for 18 subjects (six gorillas, five chimpanzees, and seven Japanese macaques). The study evaluates the efficacy of using touchscreens to test zoo-housed primates food preferences and evaluate the understanding of the photographic stimuli.

A frequentist Bradley-Terry model was used to analyze food preference. The model was fitted with the \texttt{prefmod} \citep{hatzinger2012prefmod} and the \texttt{gnm} package \citep{gnm} in R. The output of the analyses is the worth value of parameters. The analyses investigated species and subjects separately. 

This reanalysis investigates a simple Bayesian Bradley-Terry model for each species without considering the multiple judgment sampling. Then, a simple model with random effects to model subject-specific preferences is performed. A total of six models were fitted (three simple Bradley-Terry and three Bradley-Terry with random effects). Table \ref{tab:r2-waic} shows the WAIC values for each model. This table indicates that all models with random effects perform better than the models without random effects since they have lower WAICs.

Similar to the previous reanalysis, the priors were chosen to be normal distributions centered around 0 and with variance of 3.0. While this prior still regularizes and makes the model identifiable it is considered weakly informative. For the random effects, the variance was also set to 3.0, which allows large variances within the same cluster (in this example the individual primate) and being weakly informative.

\begin{table}[ht]
\caption{\label{tab:r2-waic}Comparison of the WAIC of the Bradley-Terry model and the Bradley-Terry model with random effects on the subjects for each species.}
\centering
\begin{tabular}[t]{lrr}
\toprule
\multicolumn{1}{c}{ } & \multicolumn{2}{c}{WAIC} \\
\cmidrule(l{3pt}r{3pt}){2-3}
Specie & Bradley-Terry & Bradley-Terry with random effects\\
\midrule
Macaques & 7101.5 & 6712.6\\
Chimpanzees & 7199.4 & 6720.0\\
Gorillas & 9767.2 & 8792.9\\
\bottomrule
\end{tabular}
\end{table}

\begin{figure}
\centering
\includegraphics[width=\textwidth]{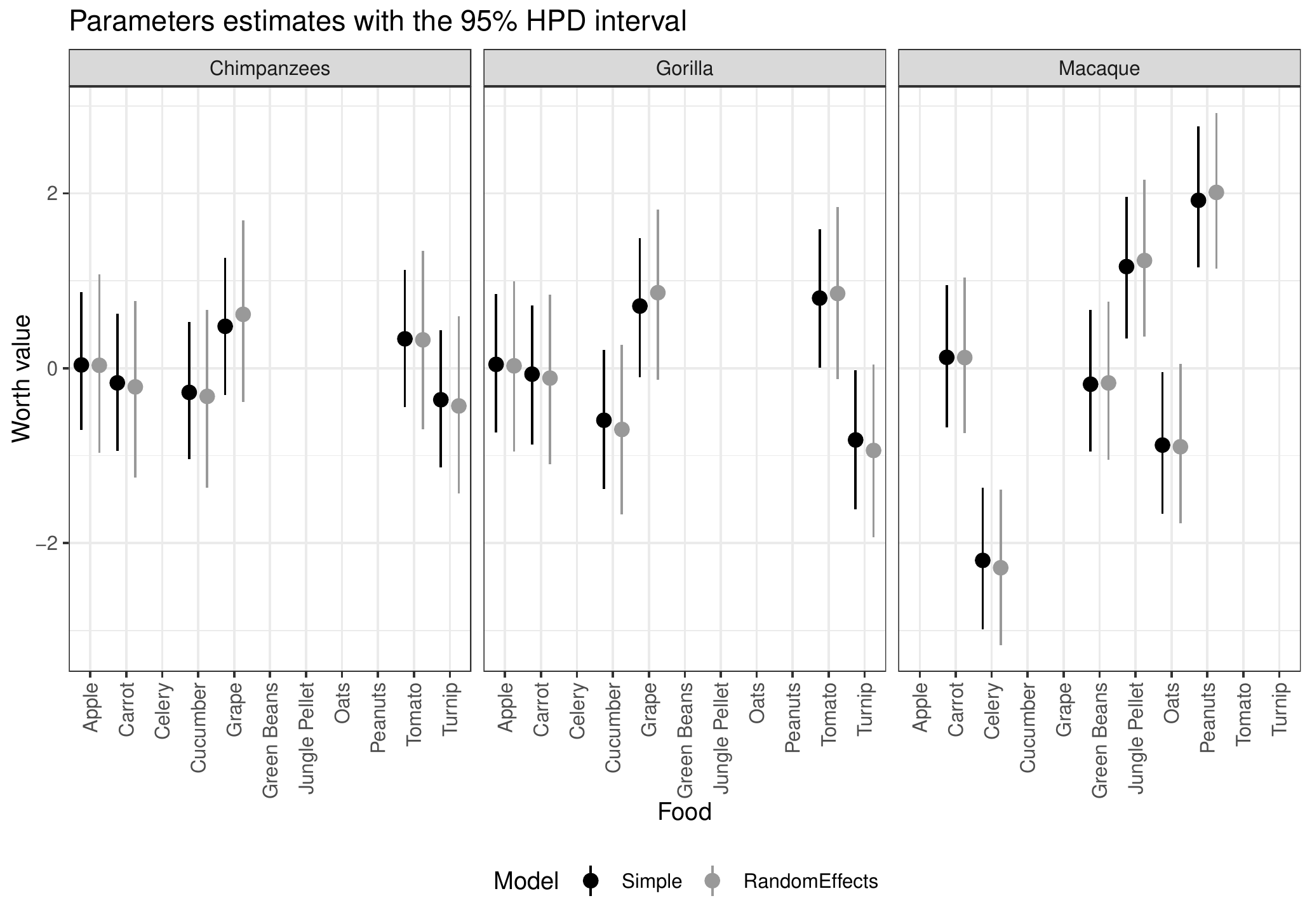}
\caption{\label{fig:r2-plot} The estimated abilities of each food type for each species in both models. Food items that do not have an estimated ability were not fed to that particular species.}
\end{figure}

Table \ref{tab:r2-parameters} shows the obtained parameters for the random effects models together with HPD intervals. 
To complement, Figure \ref{fig:r2-plot} shows a comparison of the estimates from the model with and without the random effects.  Both models show a relatively close estimated value of the abilities of each food. Without considering the random effects term, the parameter value is equivalent to analysing an average value for all individuals in the same cluster.  

However, the effect sizes represented by the actual probability of choosing one food over the other for the species and the subjects can still be different. For example, the model with random effects can estimate the probability of each subject selecting one food over the other, while the Bradley-Terry model without random effects only estimates the species average. Additionally, the random-effects model can also compensate for non-balanced data, if one subject or species has more trials than another. 

\begin{table}[ht]
\caption{\label{tab:r2-parameters}Parameters of the random effects model with 95\% HPD and the number of effective samples.}
\centering
\begin{tabular}[t]{lrrrrr}
\toprule
Parameter & Mean & Median & HPD lower & HPD upper & N. Eff. Samples\\
\midrule
\addlinespace[0.3em]
\multicolumn{6}{l}{\textbf{Macaque}}\\
\hspace{1em}Carrot & 0.12 & 0.12 & -0.74 & 1.04 & 9934\\
\hspace{1em}Celery & -2.28 & -2.29 & -3.17 & -1.39 & 9712\\
\hspace{1em}Jungle Pellet & 1.23 & 1.23 & 0.36 & 2.15 & 9778\\
\hspace{1em}Oats & -0.90 & -0.90 & -1.77 & 0.05 & 9886\\
\hspace{1em}Peanuts & 2.01 & 2.00 & 1.14 & 2.92 & 10195\\
\hspace{1em}Green Beans & -0.17 & -0.17 & -1.05 & 0.76 & 10176\\
\addlinespace[0.3em]
\multicolumn{6}{l}{\textbf{Chimpanzees}}\\
\hspace{1em}U1\_std & 0.58 & 0.57 & 0.42 & 0.75 & 4277\\
\hspace{1em}Apple & 0.03 & 0.03 & -0.97 & 1.08 & 10200\\
\hspace{1em}Tomato & 0.32 & 0.32 & -0.70 & 1.34 & 9880\\
\hspace{1em}Carrot & -0.21 & -0.21 & -1.25 & 0.77 & 9435\\
\hspace{1em}Grape & 0.62 & 0.62 & -0.38 & 1.69 & 10186\\
\hspace{1em}Cucumber & -0.32 & -0.33 & -1.36 & 0.67 & 9911\\
\hspace{1em}Turnip & -0.43 & -0.43 & -1.43 & 0.59 & 9342\\
\addlinespace[0.3em]
\multicolumn{6}{l}{\textbf{Gorilla}}\\
\hspace{1em}U1\_std & 0.72 & 0.70 & 0.49 & 1.01 & 4675\\
\hspace{1em}Apple & 0.03 & 0.03 & -0.95 & 0.99 & 13186\\
\hspace{1em}Carrot & -0.11 & -0.11 & -1.10 & 0.84 & 12365\\
\hspace{1em}Grape & 0.86 & 0.87 & -0.13 & 1.81 & 13237\\
\hspace{1em}Tomato & 0.85 & 0.86 & -0.12 & 1.85 & 13031\\
\hspace{1em}Cucumber & -0.70 & -0.70 & -1.67 & 0.27 & 12674\\
\hspace{1em}Turnip & -0.94 & -0.94 & -1.93 & 0.04 & 12944\\
U1\_std & 0.78 & 0.77 & 0.57 & 1.02 & 4883\\
\bottomrule
\end{tabular}
\end{table}

Two techniques can be used to assess the food preference. The first is through the posterior distribution of ranks of the foods. The second is through sampling the posterior distribution and calculating the probability of one stimulus being chosen when compared to another. The second method represents a measure to assess the effect size of two competing stimuli.  

Table \ref{tab:r2-rank} shows the rank of the food preferences for each species, the median, mean and standard deviations. This rank is calculated from the posterior distribution of the ability parameters. This table indicates that the macaques have a well-defined rank for peanuts, jungle pellets, oats, and celery (given the low standard deviation of the rank). However, they do not have strong preferences between carrots and green beans. Chimpanzees have less consistent ranks and with higher standard deviation. They have a higher preference for grapes and a lower preference for turnip.  Gorillas show a stronger preference for both grapes and tomatoes and a lower preference for turnips. The standard deviations on the ranks are smaller than with the chimpanzees that had the same food choice.  It is worth noting, that this analysis consist of observing the preferences at the species level, not at the individual level. For example, although chimpanzees (at the species level) did not have well-defined ranks, each subject can have defined preferences, if analyzed individually.

\begin{table}[ht]

\caption{\label{tab:r2-rank}Ranking of the food preferences per specie for the random effects model.}
\centering
\begin{tabular}[t]{lrrr}
\toprule
Food & Median Rank & Mean Rank & Std. Rank\\
\midrule
\addlinespace[0.3em]
\multicolumn{4}{l}{\textbf{Macaque}}\\
\hspace{1em}Peanuts & 1 & 1.01 & 0.09\\
\hspace{1em}Jungle Pellet & 2 & 1.99 & 0.09\\
\hspace{1em}Carrot & 3 & 3.17 & 0.37\\
\hspace{1em}Green Beans & 4 & 3.84 & 0.39\\
\hspace{1em}Oats & 5 & 4.99 & 0.08\\
\hspace{1em}Celery & 6 & 6.00 & 0.00\\
\addlinespace[0.3em]
\multicolumn{4}{l}{\textbf{Chimpanzees}}\\
\hspace{1em}Grape & 1 & 1.50 & 0.86\\
\hspace{1em}Tomato & 2 & 2.24 & 1.10\\
\hspace{1em}Apple & 3 & 3.36 & 1.25\\
\hspace{1em}Carrot & 4 & 4.26 & 1.25\\
\hspace{1em}Cucumber & 5 & 4.62 & 1.24\\
\hspace{1em}Turnip & 5 & 5.01 & 1.10\\
\addlinespace[0.3em]
\multicolumn{4}{l}{\textbf{Gorilla}}\\
\hspace{1em}Tomato & 1 & 1.54 & 0.60\\
\hspace{1em}Grape & 2 & 1.57 & 0.61\\
\hspace{1em}Apple & 3 & 3.37 & 0.68\\
\hspace{1em}Carrot & 4 & 3.69 & 0.70\\
\hspace{1em}Cucumber & 5 & 5.19 & 0.65\\
\hspace{1em}Turnip & 6 & 5.65 & 0.57\\
\bottomrule
\end{tabular}
\end{table}

The second method to assess the food preference is with the posterior probability of selecting a stimulus over the other. The \texttt{bpcs} package provides functions to calculate these probabilities for all combinations except in the case of the random effect (in which the number of combinations is much larger). However, the package also offers the possibility to calculate the probability for selected cases.

In the case of the random-effects model, it is necessary to calculate the posterior distribution of each desired pair of stimuli for each subject. However, the package has also the capability to calculate the probabilities for the average of the subjects (the case in which random effects have a null effect in the probability). Table \ref{tab:r2-probability} shows the probabilities of selecting novel stimuli against the selection of old stimuli. 

\begin{table}[ht]
\caption{\label{tab:r2-probability}Posterior probabilities of the novel stimuli i being selected over the trained stimuli j}
\centering
\begin{tabular}[t]{llrr}
\toprule
Item i & Item j & Probability & Odds Ratio\\
\midrule
\addlinespace[0.3em]
\multicolumn{4}{l}{\textbf{Gorilla}}\\
\hspace{1em}Apple & Cucumber & 0.61 & 1.56\\
\hspace{1em}Apple & Grape & 0.23 & 0.30\\
\hspace{1em}Apple & Turnip & 0.67 & 2.03\\
\hspace{1em}Apple & Carrot & 0.56 & 1.27\\
\hspace{1em}Tomato & Cucumber & 0.74 & 2.85\\
\hspace{1em}Tomato & Grape & 0.50 & 1.00\\
\hspace{1em}Tomato & Turnip & 0.87 & 6.69\\
\hspace{1em}Tomato & Carrot & 0.75 & 3.00\\
\addlinespace[0.3em]
\multicolumn{4}{l}{\textbf{Chimpanzee}}\\
\hspace{1em}Apple & Cucumber & 0.53 & 1.13\\
\hspace{1em}Apple & Grape & 0.43 & 0.75\\
\hspace{1em}Apple & Turnip & 0.54 & 1.17\\
\hspace{1em}Apple & Carrot & 0.52 & 1.08\\
\hspace{1em}Tomato & Cucumber & 0.64 & 1.78\\
\hspace{1em}Tomato & Grape & 0.41 & 0.69\\
\hspace{1em}Tomato & Turnip & 0.64 & 1.78\\
\hspace{1em}Tomato & Carrot & 0.65 & 1.86\\
\addlinespace[0.3em]
\multicolumn{4}{l}{\textbf{Macaque}}\\
\hspace{1em}Oats & Celery & 0.79 & 3.76\\
\hspace{1em}Oats & Jungle Pellet & 0.16 & 0.19\\
\hspace{1em}Oats & Peanuts & 0.05 & 0.05\\
\hspace{1em}Oats & Carrot & 0.25 & 0.33\\
\hspace{1em}Green Beans & Celery & 0.89 & 8.09\\
\hspace{1em}Green Beans & Jungle Pellet & 0.19 & 0.23\\
\hspace{1em}Green Beans & Peanuts & 0.09 & 0.10\\
\hspace{1em}Green Beans & Carrot & 0.42 & 0.72\\
\bottomrule
\end{tabular}
\end{table}

\subsection{Study III: Patients’ health locus of control and preferences about the role that they want to play in the medical decision-making process}

This reanalysis is from the article ``Patients’ health locus of control and preferences about the role that they want to play in the medical decision-making process'' \citep{marton2020patients}. In the paper, the authors investigated how Health Locus of Control (HLOC) may influence patient's preferences for active or passive roles regarding their medical decision-making. 

The study was conducted with 153 participants which responded to the Multidimensional Health Locus of Control Scale - form C \citep{ross2015multidimensional} and a series of ten paired comparisons questions. The HLOC measured four dimensions (internal, chance, doctor, and other people) using an 18-point scale. The paired comparison questions were based on hypothetical situations of the Control Preference Scale - CPS  \citep{solari2013role} in which the participants chose one scenario among a set of comparative scenarios from Active role (``I prefer to make the decision about which treatment I will receive''); Active-Collaborative role (``I prefer to make the final decision about my treatment after seriously considering my doctor’s opinion''); Collaborative role (``I prefer that my doctor and I share responsibility for deciding which treatment is best for me''); Passive-Collaborative (``I prefer that my doctor makes the final decision about which treatment will be used, but seriously considers my opinion''); or Passive role (``I prefer to leave all decisions regarding treatment to my doctor''). 

The data were analyzed with the frequentist Bradley-Terry model utilizing the \texttt{prefmod} package \citep{hatzinger2012prefmod}. Four independent models were analyzed with each dimension of HLOC as the predictor, based on median-splitting to represent high HLOC and low HLOC for each dimension. The authors opted for this approach because the \texttt{prefmod} package only supports categorical predictors. 
This reanalysis evaluates three models in increasing complexity, with the four dimensions of HLOC modeled together. The models' fits are evaluated and how it impacts the estimated coefficients. The HLOC dimensions are normalized to both be presented at a comparable scale and to facilitate inference. Centering and scaling procedures such as normalizing facilitates the convergence of predictors coefficients \cite{mcelreath2020statistical}.

The first model is a simple Bradley-Terry model, to serve as a basis. This model has a WAIC of 1422.6. The second model utilizes the four dimensions of the HLOC as predictors and has a WAIC of 1378.7. The third model introduces both random effects to compensate for individual preferences for each of the 5 roles (active to passive) and the four HLOC dimensions as subject-specific predictors. This third model has a WAIC of 801.8 indicating the best fit out of the three models. 

Similar to the previous reanalysis, the priors were chosen to be normal distributions centered around 0 and with variance of 3.0. While this prior still regularizes and makes the model identifiable it is considered weakly informative. For the random effects, the variance was also set to 3.0, which allows large variances within the same cluster and being weakly informative. For the subject specific predictors, the prior is also considered weakly informative and with a variance of 3.0.

Table \ref{tab:r3-parameters-lambda} shows the values of the obtained $\lambda$ parameters and the standard deviation of the random effects. The results indicated a higher base preference for the Collaborative role and a lower base preference for the Passive role. 

\begin{table}[ht]
\caption{\label{tab:r3-parameters-lambda}Lambda parameters of the model and the random effects standard deviation.}
\centering
\begin{tabular}[t]{lrrrrr}
\toprule
Parameter & Mean & Median & HPD lower & HPD lower & N. Eff. Samples\\
\midrule
Active & -3.16 & -3.14 & -5.98 & -0.52 & 2793\\
Active-Collaborative & 2.10 & 2.08 & -0.60 & 4.82 & 2764\\
Collaborative & 4.88 & 4.89 & 2.01 & 7.71 & 2741\\
Passive-Collaborative & 1.23 & 1.24 & -1.38 & 4.00 & 2724\\
Passive & -5.11 & -5.09 & -7.99 & -2.13 & 2718\\
U1\_std & 3.60 & 3.57 & 2.66 & 4.56 & 1899\\
\bottomrule
\end{tabular}
\end{table}

\begin{figure}
\centering
\includegraphics[width=\textwidth]{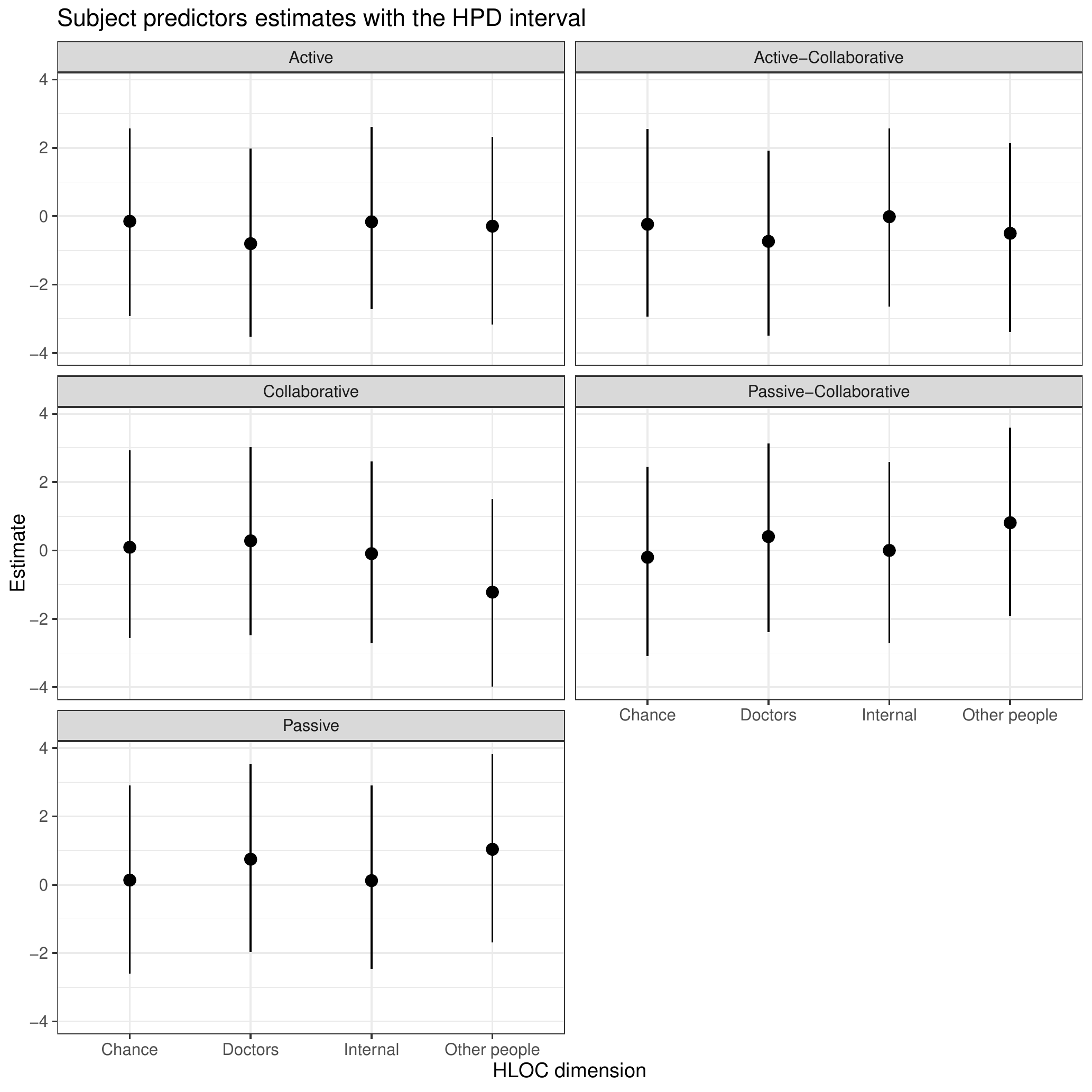}
\caption{\label{fig:r3-subject-plot} The values of the subject predictors parameters with HPD intervals.}
\end{figure}

Table \ref{tab:r3-parameters-subject} shows the parameters of the subject predictors. This table shows the values of the subject predictors parameters by each type of role. This table is more easily visualized with a plot, as shown in Figure \ref{fig:r3-subject-plot}. Table \ref{tab:r3-parameters-subject} and figure \ref{fig:r3-subject-plot} show the uncertainty in the actual impact of the HLOC in the CPS role. Most estimates have a median value close to zero and large HPD intervals overlapping zero. A similar conclusion can also be assessed in the probabilities of a selecting a specific CPS role, as shown in table  \ref{tab:r3-probability}.  

Table \ref{tab:r3-probability} shows a few cases illustrating the impact of the HLOC dimensions in the actual probability of selecting a specific CPS role. Specifically, this table shows the probabilities of a subject to select between the roles Active and Passive, Active-Collaborative and Collaborative, and Collaborative and Passive-Collaborative with changes of the HLOC in the different dimensions. This table shows that the probability choice between the roles active and passive for the mean value is 0.88. For a subject with internal HLOC two standard deviations below the average this probability changes to 0.80, when the internal HLOC is two standard deviations above this probability in unchanged compared to the average. This indicates that although internal HLOC has a small impact in selecting between these two roles. These small changes in probabilities indicate that while using HLOC dimensions as subject predictors improves the model, their effect on the actual probabilities of selecting a role are small. The baseline coefficients for each role contribute more relative difference between the roles than the effect of the subject-specific predictors.

\begin{table}[ht]

\caption{\label{tab:r3-parameters-subject}Subject predictors parameters by role.}
\centering
\begin{tabular}[t]{lrrrrr}
\toprule
Parameter & Mean & Median & HPD lower & HPD lower & N. Eff. Samples\\
\midrule
\addlinespace[0.3em]
\multicolumn{6}{l}{\textbf{Active}}\\
\hspace{1em}Internal & -0.16 & -0.16 & -2.72 & 2.61 & 2563\\
\hspace{1em}Chance & -0.15 & -0.15 & -2.93 & 2.56 & 2494\\
\hspace{1em}Doctors & -0.80 & -0.80 & -3.53 & 1.97 & 2211\\
\hspace{1em}Other people & -0.29 & -0.28 & -3.16 & 2.33 & 2576\\
\addlinespace[0.3em]
\multicolumn{6}{l}{\textbf{Active-Collaborative}}\\
\hspace{1em}Internal & -0.01 & -0.01 & -2.64 & 2.56 & 2534\\
\hspace{1em}Chance & -0.24 & -0.25 & -2.94 & 2.55 & 2374\\
\hspace{1em}Doctors & -0.74 & -0.73 & -3.50 & 1.92 & 2204\\
\hspace{1em}Other people & -0.50 & -0.50 & -3.39 & 2.14 & 2594\\
\addlinespace[0.3em]
\multicolumn{6}{l}{\textbf{Collaborative}}\\
\hspace{1em}Internal & -0.09 & -0.08 & -2.72 & 2.61 & 2571\\
\hspace{1em}Chance & 0.09 & 0.09 & -2.56 & 2.93 & 2362\\
\hspace{1em}Doctors & 0.28 & 0.29 & -2.48 & 3.02 & 2190\\
\hspace{1em}Other people & -1.22 & -1.23 & -3.98 & 1.51 & 2703\\
\addlinespace[0.3em]
\multicolumn{6}{l}{\textbf{Passive-Collaborative}}\\
\hspace{1em}Internal & 0.12 & 0.12 & -2.46 & 2.90 & 2516\\
\hspace{1em}Chance & 0.13 & 0.14 & -2.60 & 2.90 & 2355\\
\hspace{1em}Doctors & 0.75 & 0.76 & -1.97 & 3.54 & 2174\\
\hspace{1em}Other people & 1.04 & 1.04 & -1.69 & 3.81 & 2708\\
\addlinespace[0.3em]
\multicolumn{6}{l}{\textbf{Passive}}\\
\hspace{1em}Internal & 0.00 & 0.01 & -2.71 & 2.59 & 2509\\
\hspace{1em}Chance & -0.20 & -0.20 & -3.08 & 2.45 & 2375\\
\hspace{1em}Doctors & 0.41 & 0.40 & -2.38 & 3.12 & 2205\\
\hspace{1em}Other people & 0.81 & 0.81 & -1.90 & 3.59 & 2656\\
\bottomrule
\end{tabular}
\end{table}

\begin{table}[ht]
\caption{\label{tab:r3-probability}Probabilities of selecting role i instead of j based on changes of the values of the HLOC dimensions}
\centering
\begin{tabular}[t]{llrrrrr}
\toprule
\multicolumn{2}{c}{Roles} & \multicolumn{4}{c}{HLOC dimensions} & \multicolumn{1}{c}{ } \\
\cmidrule(l{3pt}r{3pt}){1-2} \cmidrule(l{3pt}r{3pt}){3-6}
i & j & Internal & Chance & Doctors & OtherPeople & Probability\\
\midrule
Active & Passive & 0 & 0 & 0 & 0 & 0.88\\
Active & Passive & -2 & 0 & 0 & 0 & 0.80\\
Active & Passive & 2 & 0 & 0 & 0 & 0.88\\
Active & Passive & 0 & 0 & -2 & 0 & 0.89\\
Active & Passive & 0 & 0 & 2 & 0 & 0.78\\
\addlinespace
Act.-Collab. & Collab. & 0 & 0 & 0 & 0 & 0.08\\
Act.-Collab. & Collab. & -2 & 0 & 0 & 0 & 0.08\\
Act.-Collab. & Collab. & 2 & 0 & 0 & 0 & 0.07\\
Act.-Collab. & Collab. & 0 & 0 & -2 & 0 & 0.07\\
Act.-Collab. & Collab. & 0 & 0 & 2 & 0 & 0.04\\
\addlinespace
Collab. & Pass.-Collab. & 0 & 0 & 0 & 0 & 0.98\\
Collab. & Pass.-Collab. & 0 & -2 & 0 & 0 & 0.94\\
Collab. & Pass.-Collab. & 0 & 2 & 0 & 0 & 0.98\\
Collab. & Pass.-Collab. & 0 & 0 & 0 & -2 & 0.97\\
Collab. & Pass.-Collab. & 0 & 0 & 0 & 2 & 0.96\\
\bottomrule
\end{tabular}
\end{table}

\section{Conclusion}
The ultimate goal of this article is to provide tools with strong theoretical foundations that empower researchers to have alternatives to the use of frequentist data analysis when analyzing paired data. Therefore, the \texttt{bpcs} package was introduced to facilitate the adoption of Bayesian models in paired comparison assessments. The package is free to use, and the latest version is available at the official repository. 

This article explained the rationales behind the different Bayesian models implemented in the \texttt{bpcs} package. Additionally, the article provides the reanalyses of three studies in various areas of behavior science (psychophysics, animal research, and health). The package allows researchers to run the Bayesian Bradley-Terry model and many of its extensions, such as the Davidson model to handle ties, models with order effect, generalized models, models with dependent data, and models with predictors on the subject (and the different combinations of these extensions). It also provides tools for assessing uncertainty in the ranks and the posterior probabilities, not available in frequentist packages. All the code used to fit the models, create the tables and the figures from the reanalyses section are available in the online appendix.

Being able to easily extend a simple model to more complex ones, as shown in the reanalyses,  allows researchers to control bias and errors in the modeling. Future research could further develop Bayesian cumulative models (when there is a strength scale in the assessment of two items) and models with time dependency.

\bibliography{bibliography}

\end{document}